\documentclass[letterpaper]{article} 
\usepackage{amssymb}

\usepackage{aaai2026}  
\usepackage{times}  
\usepackage{helvet}  
\usepackage{courier}  
\usepackage[hyphens]{url}  
\usepackage{graphicx} 
\usepackage{caption}
\usepackage{subcaption}
\usepackage{natbib}  
\usepackage{caption} 
\usepackage{algorithm}
\usepackage{algorithmic}
\usepackage{newfloat}
\usepackage{listings}
\usepackage{amsmath}
\usepackage{xcolor}
\usepackage[most]{tcolorbox}
\usepackage{longtable}

\nocopyright

\DeclareCaptionStyle{ruled}{labelfont=normalfont,labelsep=colon,strut=off} 
\floatstyle{ruled}
\newfloat{listing}{tb}{lst}{}
\floatname{listing}{Listing}
\title{Interpreting Multimodal Communication at Scale in Short-Form Video: \\\
Visual, Audio, and Textual Mental Health Discourse on TikTok}
\author{
    Mingyue Zha\textsuperscript{\rm 1}, 
    Ho-Chun Herbert Chang\textsuperscript{\rm 1}
}

\affiliations{
    \textsuperscript{\rm 1} Program in Quantitative Social Science, Dartmouth College, Hanover, NH 03755 USA \\

}

\usepackage{bibentry}

\begin{document}

\maketitle

\begin{abstract}
Short-form video platforms integrate text, visuals, and audio into complex communicative acts, yet existing research analyzes these modalities in isolation, lacking scalable frameworks to interpret their joint contributions. This study introduces a pipeline combining automated multimodal feature extraction with Shapley value-based interpretability to analyze how text, visuals, and audio jointly influence engagement. Applying this framework to 162,965 TikTok videos and 814,825 images about social anxiety disorder (SAD), we find that facial expressions outperform textual sentiment in predicting viewership, informational content drives more attention than emotional support, and cross-modal synergies exhibit threshold-dependent effects. These findings demonstrate how multimodal analysis reveals interaction patterns invisible to single-modality approaches. Methodologically, we contribute a reproducible framework for interpretable multimodal research applicable across domains; substantively, we advance understanding of mental health communication in algorithmically mediated environments.
\end{abstract}

\section{Introduction}

The rise of social media has fundamentally transformed interpersonal communication, allowing for digitally mediated exchanges that are networked, asynchronous, and public. The recent proliferation of multimodal platforms has integrated text, images, video, audio, and ephemeral content across synchronous and asynchronous formats. Platforms like Instagram, TikTok, and Snapchat privilege visual and performative communication over text-based exchange~\citep{violot2024shorts, huang2025parasocial, xie2025recovery}, demanding "multimodal literacy" and have reshaped how identity, emotion, and relational connection are expressed online. 
Kress and van Leeuwen conceptualize meaning-making now occurs through integrated semiotic systems rather than verbal language alone~\citep{kress2001multimodal}. Online, interpersonal communication now operate at the milieu of visual, textual, and algorithmic, with necessitate not just a theoretical, but methodological framework, that account for the various sensory dimensions of digital interaction~\citep{andalibi2017sensitive, niu2021stayhome}.

Despite the growing importance of short-form videos, methodological tools for analyzing such content at scale remain underdeveloped. Much existing social media research focuses on single modalities (most commonly text)~\citep{islam2025towards}, while multimodal studies often rely on small samples and manual annotation~\citep{wang2021multimodal, wu2025hotbed}. As a result, there is limited consensus on how to systematically extract and interpret multimodal signals from a large dataset of short-form videos~\citep{ali2025systematic}. Researchers need a principled workflow for comparing the relative contributions of textual, visual, and audio features within a unified framework.

Recent advances in machine learning, including pretrained language models, image recognition, and audio embeddings, make such integration computationally feasible at scale~\citep{wang2023large}. Emerging studies demonstrate that LLM-assisted content analysis can effectively extract and annotate visual features from online videos ~\citep{li2024understanding, liu2025can}. However, these tools are often deployed with less attention to interpretability, reproducibility, or cross-modal comparability. 

This paper addresses this gap by introducing a scalable pipeline for multimodal analysis of short-form social media videos. Our approach integrates automated feature extraction across text, image, and audio modalities with Shapley value–based interpretability, enabling direct comparison of heterogeneous features within a common explanatory framework. By treating multimodal annotations as probabilistic inputs and applying Shapley value analysis, our method attributes attention and engagement outcomes to not just individual modal features but their interaction. Additionally, we build an interpretable framework on top of point-level explainers to aid social scientific interpretation. 

We demonstrate the utility of this methodology through a case study of mental health discourse on social media. Mental health content provides a particularly compelling case for multimodal analysis, as affective expression is distributed across written, spoken, and visual presentation~\citep{yazdavar2020multimodal}. In this paper, we examine how different multimodal elements shape attention dynamics in mental health TikToks related to social anxiety disorder (SAD). 


\section{Literature Review}

\subsection{Multimodal Communication in Digital Spaces}

The shift from text-dominant to multimodal social media platforms represents a fundamental transformation in how communication occurs online. Kress and van Leeuwen argue that contemporary communication operates through integrated semiotic systems rather than verbal language alone, where visual, textual, and auditory elements function as distinct but interconnected meaning-making resources \cite{kress2001multimodal}. In digital contexts, these modalities interact, sometimes reinforcing and sometimes contradicting one another.

Short-form video platforms like TikTok, Instagram Reels, and YouTube Shorts exemplify this multimodal complexity. These platforms privilege visual performance and algorithmic curation over traditional text-based discourse. The affordances of these platforms—vertical video format, music integration, text overlay capabilities, rapid content consumption—create unique communicative environments where success depends on coordinating elements across modalities. Media richness theory suggests that face-to-face communication is traditionally considered the ``richest'' medium due to its integration of verbal, vocal, and visual cues \cite{daft1986organizational}. Short-form video platforms approximate this richness in asynchronous, algorithmically mediated formats.

Yet we know surprisingly little about how these multimodal elements function in practice, and integrated frameworks to conduct these analyses are growing but still sparse.

\subsection{Mental Health Disclosure in Digital Spaces}

Social media has become a primary site for mental health communication, peer support, and health information seeking. Research on mental health disclosure online reveals complex motivations: individuals seek emotional validation, practical coping strategies, reduced stigma through normalized discussion, and connections with others sharing similar experiences \cite{andalibi2017sensitive}. Platforms like TikTok have emerged as particularly important spaces for youth mental health discourse, with users creating content that ranges from personal narrative and peer support to psychoeducation and advocacy \cite{milton2023see}.

Self-disclosure theory suggests that revealing personal information serves relationship-building functions, but digital self-disclosure operates under different constraints than face-to-face interaction. Online disclosures are often public, persistent, and algorithmically mediated, reaching audiences that extend far beyond immediate social networks \cite{bazarova2014self}. The permanence and visibility of digital disclosures can amplify both benefits (finding community, reducing isolation) and risks (privacy concerns, unwanted attention, stigmatization). For individuals with social anxiety disorder specifically—characterized by intense fear of social evaluation and performance situations—the act of creating video content that foregrounds the self presents a notable paradox. Visual self-presentation, the core anxiety trigger for those with SAD, becomes the primary medium through which individuals discuss their social anxiety experiences.

Studies of online support communities identify different forms of social support: emotional support (expressions of empathy and caring), informational support (advice and practical guidance), instrumental support (tangible assistance), and appraisal support (feedback for self-evaluation) \cite{cutrona1992controllability}. These distinctions matter because different types of content may serve different audience needs and may be differentially valued by platform algorithms.

Recent work examining TikTok mental health content specifically has found that algorithmic recommendation systems shape which mental health narratives become visible, with potential implications for both helpful peer support and misinformation spread \cite{bickham2024hidden}. Milton et al.~\cite{milton2023see} document how TikTok users with mental health conditions find community and validation through algorithmically curated content, but also note concerns about the quality and accuracy of peer-generated health information. Social media conversations around social anxiety thus become a useful testbed, due to the hyper self-aware nature of interpersonal communication and self-presentation. However, prior studies largely relied on qualitative methods or small-scale quantitative analysis unable to capture the full complexity and diversity of multimodal expression.

\subsection{Multimodal Analysis of Social Media: Current Approaches and Limitations}

Research on social media viewership has largely proceeded along single-modality lines. Text-based studies examine how linguistic features—sentiment, lexical diversity, emotional intensity, linguistic style—predict popularity outcomes \cite{vilares2015usefulness, hong2013co}. Visual analysis focuses on image aesthetics, color composition, facial presence, and object recognition \cite{trzcinskipredicting, wu2017sequential}. Audio studies, though less common, investigate music characteristics, vocal prosody, and speech patterns \cite{li2024understanding}. These single-modality approaches have generated valuable insights but necessarily provide incomplete accounts of how multimodal content functions.

The few existing multimodal studies typically combine features from two modalities—most commonly text and images—and feed concatenated feature vectors into prediction models \cite{meghawat2018multimodal}. While this approach can improve predictive accuracy over single-modality models, it treats multimodal features as a flat input space, making it difficult to understand how different modalities individually contribute or interact. More critically, these studies often rely on small samples with manual annotation, limiting both the scale of analysis and the generalizability of findings. As \citet{chen2025labeling} demonstrate, multimodal annotation requires substantial human labor, and the rapid evolution of platform norms means that manually coded datasets quickly become outdated.

Recent work in computer science uses deep learning to analyze how visual, audio, and textual features jointly predict TikTok creator influence tiers. However, as is common with deep learning, embeddings sacrifice interpretability—researchers cannot easily identify which specific features drive predictions or how features from different modalities interact. For social scientists interested in understanding the mechanisms through which content characteristics shape attention, prediction accuracy alone is insufficient.

\subsection{Interpretability in Computational Social Science}

For social scientists, interpretability is not merely a technical desideratum—it is fundamental to scientific generalizability and normatively, the design of interventions. Shapley values, derived from cooperative game theory, offer a principled approach to model interpretation \cite{winter2002shapley}. The core insight is to treat each feature as a ``player'' contributing to a prediction ``payout,'' and to compute each feature's contribution by considering its average marginal effect across all possible coalitions of other features. SHAP (SHapley Additive exPlanations) operationalizes this concept for machine learning models, providing a unified framework that decomposes any prediction into additive feature contributions \cite{lundberg2017unified}.

Recent applications of SHAP in computational social science have demonstrated its value for understanding feature importance in tabular data settings \cite{salih2025perspective}. However, SHAP has not been systematically applied to multimodal social media analysis in ways that preserve and leverage modality structure. Standard global SHAP summaries (e.g., mean absolute attribution) exist, but they do not provide a coefficient-like directional summary conditioned on feature presence/intensity, which is often what social scientists want.

The existing literature reveals several critical gaps. First, little prior work has worked to develope a comprehensive automated pipeline for extracting theory-driven features across text, visual, and audio modalities from short-form videos at scale. Second, while multimodal prediction models exist, they often lack the interpretability needed to understand how specific features within and across modalities shape outcomes. Third, mental health communication research has not systematically examined how affective expression diverges across modalities.

This study addresses these gaps by introducing an automated multimodal analysis pipeline that integrates zero-shot classification for semantic features with Shapley value-based interpretability. We apply this framework to mental health discourse about social anxiety disorder on TikTok, examining three research questions:

\textbf{RQ1}: How do textual, audio, and visual sentiment diverge in characterizing mental health-related discourse?

\textbf{RQ2}: To what extent do textual, visual, and audio modalities contribute to attention dynamics in short-form mental health-related videos?

\textbf{RQ3}: How do textual, visual, and audio features jointly shape attention dynamics in short-form mental health videos?

By answering these questions, we contribute both methodologically (a reproducible framework for interpretable multimodal analysis) and substantively (new insights into how mental health communication functions in algorithmically mediated environments).

\section{Methodology}

\subsection{Overview of Multimodal Pipeline}
Our analytical pipeline integrates textual, visual, and audio information from social media videos into a unified, interpretable multimodal framework. Figure 1 provides a schematic overview of the pipeline. Each video is decomposed into three modalities, which are processed independently using modality-appropriate feature extraction methods. Textual content is analyzed using sentiment analysis and zero-shot discourse labeling, visual frames are coded using facial analysis and zero-shot visual labeling, and audio tracks are analyzed for acoustic features with spoken content transcribed for downstream textual analysis. Across modalities, zero-shot labeling produces probabilistic feature representations that enable additive and comparable attribution across heterogeneous feature types. All features are concatenated into a unified multimodal model, and Shapley value analysis is applied to quantify feature- and modality-level contributions to model predictions.

\begin{figure}[!htb]
    \includegraphics[width=0.5\textwidth]{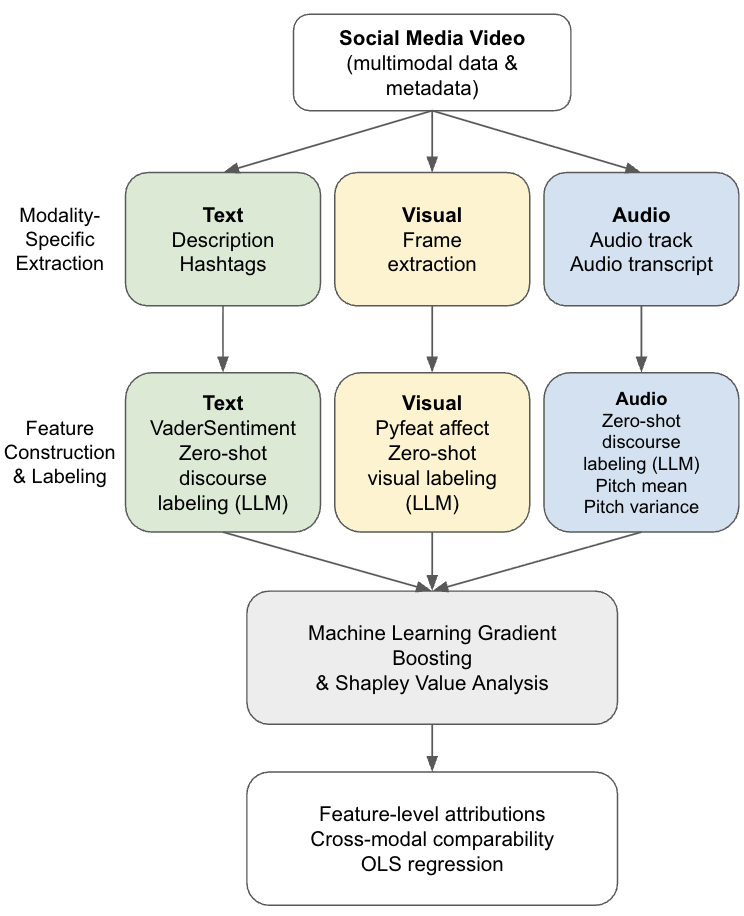}
    \caption{Overview of the multimodal analysis pipeline }
    \label{fig:importance}
\end{figure}

\subsection{Data extraction}

We collected data via the TikTok Research API using keywords related to social anxiety disorder: "social anxiety", "socialanxiety", "social phobia", "socialphobia", "social anxiety disorder", "socialanxietydisorder". Our final dataset included 162,965 posts published in the United States between January 1, 2020, and January 1, 2025. For each post, the API returned metadata including upload date, video URL, description text, like count, comment count, and view count. All corresponding videos were downloaded using the Python module Pyktok for multimodal analysis.

\subsection{Textual Modality }

\subsubsection{Text Extraction}
We analyzed sources of textual content associated with each video from video descriptions captions and hashtags. All textual content was minimally preprocessed through lowercasing and removal of URLs and nonsemantic artifacts, preserving informal language, emojis, and punctuation where present to retain affective and stylistic signals characteristic of social media discourse.

\subsubsection{Sentiment Analysis}
To quantify affective tone, we applied VADER (Valence Aware Dictionary and sEntiment Reasoner), a rule-based sentiment analysis tool optimized for short, informal social media text \cite{elbagir2019twitter}. VADER produced four scores for each text input: positive, negative, neutral, and compound sentiment.

\subsubsection{Zero-shot Discourse Labeling}
Beyond surface-level sentiment, we employed zero-shot classification to capture higher-order communicative intent. We used GPT-4o-Mini to assign probabilistic labels across predefined discourse categories, including self-disclosure, coping strategies, humor, informational support, emotional support, and other theoretically motivated constructs relevant to mental health communication. Video descriptions and hashtags were passed to the model using a structured prompt requesting probability scores for each category rather than discrete class assignments. Treating discourse labels as continuous probability distributions (0-100) allowed these features to integrate naturally into downstream predictive models. The full classification prompt is documented in Appendix A.


\subsection{Visual Modality}

\subsubsection{Frame Extraction}
For each video, we extracted five frames evenly spaced across the video duration. This sampling strategy was chosen to balance representational coverage of visual content with computational efficiency and consistency across videos of slightly varying lengths.

Evenly spaced sampling ensures that frames capture visual information from the beginning, middle, and end of each video, reducing bias toward temporally localized moments such as openings or climactic segments. Compared to random sampling, this approach provides standardized coverage and reproducibility~\citep{brkic2025frame}. Compared to dense frame extraction, it reduces computational cost while preserving important visual cues related to video format, subjects, and environment, which tend to be relatively stable in short-form videos. Prior work in multimodal and video-based social media analysis suggests that a small number of representative frames is sufficient for capturing high-level visual characteristics relevant to perception and engagement~\citep{kandhare2024empirical,kennedy2025demographic}. Five frames per video is a good compromise between coverage and efficiency, enabling scalable analysis.

\subsubsection{Affective Analysis}
Frames containing faces were analyzed using open-sourced PyFeat \cite{cheong2023py}, which extracts Action Units (facial muscle movements) and classifies emotions into seven categories: happiness, sadness, anger, fear, surprise, disgust, and neutral. Each emotion was quantified with intensity values ranging from 0 to 1. We aggregated these into positive affect (happiness, surprise, neutral) and negative affect (anger, disgust, fear, sadness) measures.

\subsubsection{Zero-shot Visual Labeling}
To systematically code higher-level visual characteristics, each frame was passed to GPT-4o-Mini using a structured visual prompt requesting probability scores for visual attributes including video format (selfie, professional, homemade, meme), shot type (close-up, wide, point-of-view), presence of special effects, number of people visible, and environmental context. This zero-shot visual labeling approach mirrors the textual labeling pipeline, enabling consistent annotation rigor across modalities. The complete visual classification prompt is provided in Appendix B.

\subsection{Audio Modality}

\subsubsection{Audio Extraction}
Audio tracks were extracted from video files using pydub's AudioSegment class and converted to uncompressed WAV format. Pydub is a widely-used, open-source Python library for audio manipulation~\citep{pydub_audio_python, ffmpeg_official}. These files served as inputs for audio feature extraction and analysis.  

A limitation of our audio analysis was that short-form videos, especially those on Tiktok, sometimes overlay music with speech. We acknowledge this as a measurement limitation and interpret audio features as composite signals rather than isolating speech or music independently.

\subsubsection{Acoustic Feature Extraction}
We extracted acoustic features using Aubio \cite{brossier2006automatic}, computing measures of mean pitch, pitch variability, and the presence of background music. These features capture paralinguistic and musical cues that contribute to production style.

\subsubsection{Speech Transcription and Labeling}
We employed OpenAI Whisper \cite{radford2023robust} to generate transcripts from each audio track. These transcripts were incorporated into the textual modality and processed using the same sentiment and discourse labeling pipelines described above.

\subsection{Predictive Modeling and Shapley Value Analysis}

\subsubsection{Modeling Framework}

To evaluate how multimodal features shape attention dynamics, we employed gradient boosting regression models using CatBoost \cite{prokhorenkova2018catboost}. Gradient boosting methods accommodate heterogeneous feature types, capture nonlinear relationships and higher-order interactions, and achieve strong predictive performance~\citep{chen2016xgboost}. We used logged view count as the dependent variable, as it represents initial attention capture rather than post-viewing behaviors (likes, comments, shares). All input features were normalized to a 0-100 scale to ensure standardization across variables, where 100 represents full confidence of the feature being present.

We configured the CatBoost regressor with 1,000 iterations, a learning rate of 0.1, and a maximum tree depth of 6. We trained models on an 80-20 train-test split to evaluate predictive performance.


\subsubsection{SHAP Value Analysis}

To interpret predictions from our gradient boosting models, we use SHAP (SHapley Additive exPlanations), which attributes each prediction to input features using Shapley values from cooperative game theory~\citep{lundberg2017unified}. For a fitted model $f(\mathbf{x})$ and a background (reference) distribution over inputs $\mathbf{X}$, SHAP constructs an additive explanation model
\begin{equation}
f(\mathbf{x}) \;=\; \mathbb{E}\!\left[f(\mathbf{X})\right] \;+\; \sum_{i=1}^{p} \phi_i(\mathbf{x}),
\end{equation}
where $\mathbb{E}[f(\mathbf{X})]$ is the baseline prediction and $\phi_i(\mathbf{x})$ is feature $i$'s contribution (in the units of the model output) for observation $\mathbf{x}$. 
Shapley values are defined as the average marginal contribution of a feature across all subsets (coalitions) of the remaining features:
\begin{equation}
\phi_i(\mathbf{x}) \;=\; \sum_{S \subseteq N \setminus \{i\}}
\frac{|S|!\,(|N|-|S|-1)!}{|N|!}
\left[\,v(S \cup \{i\}) - v(S)\,\right]
\end{equation}
where $N=\{1,\dots,p\}$ are index features and $v(S)$ denote the model value when only features in subset $S \subseteq N$ are present.

SHAP can be further decomposed into main effects and pairwise interaction effects. 

For each observation $\mathbf{x}$, SHAP interaction values allocate the deviation from the baseline prediction into (i) a main-effect contribution for each feature and (ii) additional terms capturing how pairs of features jointly contribute beyond their individual effects:
\begin{equation}
f(\mathbf{x}) \;=\; \mathbb{E}\!\left[f(\mathbf{X})\right] \;+\; \sum_{i=1}^{p} \phi_{ii}(\mathbf{x}) \;+\; \sum_{i<j} \phi_{ij}(\mathbf{x}).
\end{equation}
where, $\phi_{ii}(\mathbf{x})$ denotes the main-effect attribution for feature $i$, and $\phi_{ij}(\mathbf{x})$ for $i\neq j$ denotes the interaction attribution between features $i$ and $j$. In additive models without interactions between $i$ and $j$, $\phi_{ij}(\mathbf{x})$ is approximately zero across observations.

\subsection{Linearization of Estimator Surfaces}
While SHAP beeswarm plots reveal heterogeneity across individual predictions, social scientists typically require aggregate measures of feature importance analogous to regression coefficients. We address this by computing feature-weighted SHAP contributions. We first standardize the underlying features. The centering operation is domain-specific(i.e. mean or media); we elect 0.5 as our annotations represent probabilities.

Feature-weighted SHAP contributions can therefore be expressed as:
\begin{equation}\label{eq:feat-shap}
    \beta_{i,SHAP} = \sum_{k=0}^N x_k \phi_k
\end{equation}
This aggregation weights each observation's SHAP value by its feature magnitude, providing a summary measure of how strongly feature ii
i influences predictions when present. Unlike raw SHAP means, this approach accounts for feature intensity: a feature with consistently high SHAP values only when strongly present (high $x_k$) receives appropriate weight.

\subsection{Piecewise OLS on SHAP}

To characterize threshold-dependent interactions between features from different modalities, we partitioned the feature space at median values (50 for probabilistic features) and estimated separate linear relationships within each quadrant. In the continous domain, this corresponds to the SHAP value $z$ with respect to variables $x$ and $y$: 
$
z'' = \tfrac{\partial^2 z}{\partial x \partial y}
$.

Suppose there exists a median threshold $x_0$ such that the effect of $y$ diverges once $x$ exceeds this threshold:
\[
\begin{cases}
z'' > 0, \qquad \text{if } x > x_0 \quad \& \quad y > y_0, \\
z'' < 0, \qquad \text{if } x > x_0 \quad \& \quad y < y_0, \\
z'' = 0, \qquad \text{if } x < x_0 .
\end{cases}
\]

To characterize the analogous case in which the interaction is rotated along the $y$-axis, the conditions on $x$ are inverted (i.e., $z'' = 0$ when $x > x_0$). 
The third scenario arises when the direction of interaction depends on the quadrant defined by the thresholds of both variables, yielding a sign change across quadrants:
\[
\begin{cases}
z'' > 0, \qquad  \text{if } x y > 0, \\
z'' < 0, \qquad  \text{if } x y < 0 .
\end{cases}
\]
This approach reveals whether synergies emerge only when both features exceed thresholds (asymmetric interactions) or operate across all quadrants (symmetric interactions). Because SHAP values are centered at zero by default, the correlation coefficient directly estimates the linear effect of the interaction within each quadrant.


\subsection{Within-modality and cross-modal Analysis}

Using CatBoost, we computed SHAP values separately within each modality to quantify how individual textual, visual, and audio features contributed to predicted view counts. This within-modality approach allowed us to identify the most influential features while preserving the internal structure of each modality.
To enable cross-modal comparisons and assess interactions between features from different modalities, we employed XGBoost \cite{chen2016xgboost}, training models on the same 80-20 train-test split. The model was trained for up to 10,000 boosting rounds with early stopping based on test set performance, evaluated every 1,000 rounds to monitor convergence. Feature-level SHAP values were then analyzed to examine interactions between features from different modalities.

\subsection{Validation}

To ensure the accuracy of our zero-shot classifications, we manually validated a random sample of 200 frames (captions, transcripts, and images) against human annotations.  Validation results (Appendix C) demonstrate overall accuracy rates of 90\% for captions, 86.5\% for transcripts, and 83.5\% for images. Categories with clear behavioral referents (e.g., Healthcare: 100\%, Self-Disclosure: 100\%) showed near-perfect agreement, while more abstract constructs showed lower validity (e.g., Situational Stressors: 65\%, AI-generated content: 60\%). We retain all features to maintain theoretical completeness but interpret low-validity features cautiously in our results.

\section{Results}

\subsection{Modality Specific Feature Contributions}
\begin{figure}[h!]
    \includegraphics[width=0.49\textwidth]{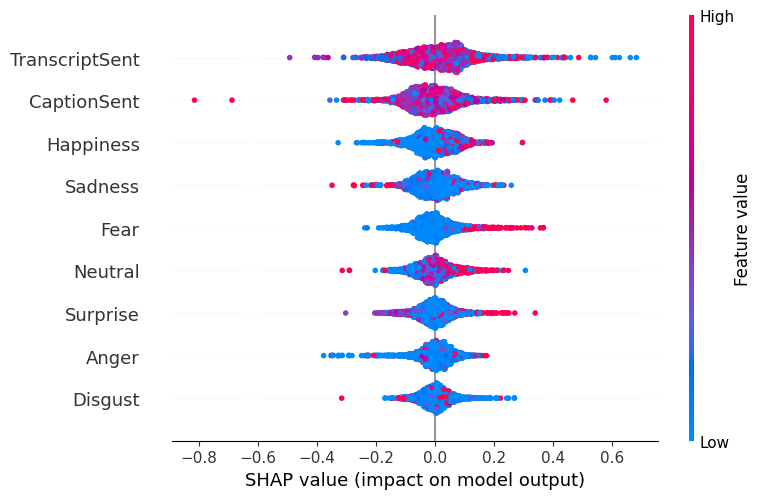}
    \includegraphics[width=0.45\textwidth]{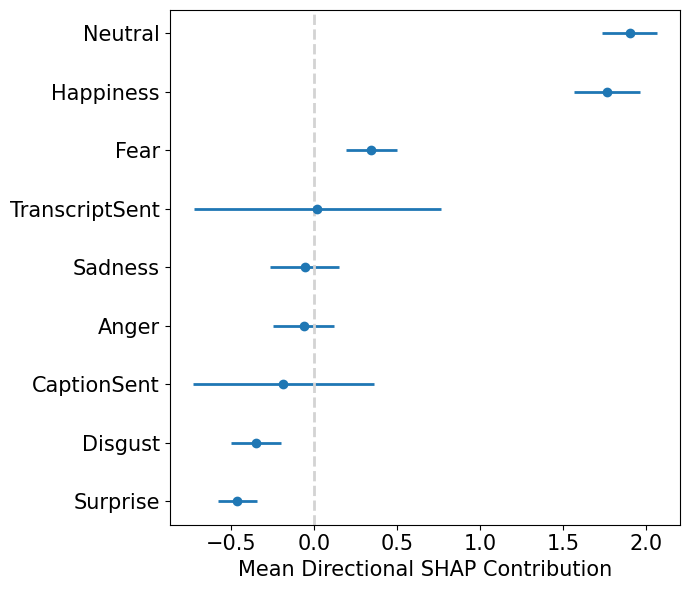}
    \caption{a) SHAP beeswarm plot of features from the visual, text, and audio modalities and b) point-weighted aggregate SHAP values as a forest plot.}
    \label{fig:shap_sentiment}
\end{figure}

First, we consider how sentiment features across modalities differentially predict exposure outcomes using SHAP values. Figure~\ref{fig:shap_sentiment}a presents a beeswarm plot of SHAP value distributions for sentiment features extracted from textual (caption), visual (facial expression), and audio (speech transcript) modalities.  The color gradient shows feature magnitude (high values in red/pink, low in blue), while horizontal position indicates impact on viewership. Facial expressions of happiness and neutral affect seems to predict higher viewership, with positive SHAP values (pink) concentrated right of zero, whereas caption and transcript sentiment measures show noisy, scattered patterns across both positive and negative ranges.

While the beeswarm plot is valuable for exploring distributional patterns, it lacks interpretability. For instance, although both transcript and caption sentiment have high feature importance, such importance is not necessarily clear. Utilizing feature-weighted SHAP (Eq.~\ref{eq:feat-shap}), direct interpretation of direction and magnitude becomes possible. Figure~\ref{fig:shap_sentiment} reveals that facial expressions of happiness and neutral affect have clear positive SHAP contributions compared to textual and speech sentiment, whose SHAP values span a large range. 
This suggests that in multimodal content, creator facial expressions produce unambiguous influence on virality~\citep{slepian2019facial, horstmann2003facial}.

\begin{figure}[H]
    \includegraphics[width=0.48\textwidth]{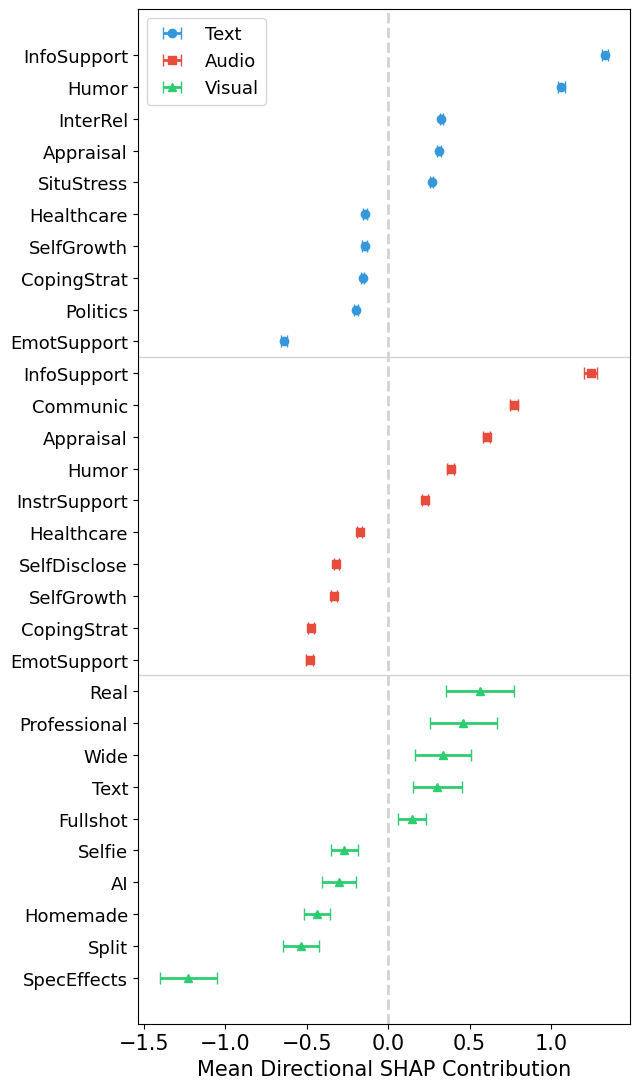}
    \caption{Mean directional SHAP contributions for textual (blue), audio (red), and visual (green) features.}
    \label{fig:big_shap}
\end{figure}

To identify which modality-specific features most strongly predict viewership more generally, we computed feature-weighted SHAP values for all textual, visual, and audio features. Figure~\ref{fig:big_shap} presents the ten most influential features from each modality (five with the clearest positive effects and five with the most negative). 

Textual and audio analysis revealed that posts labeled as relating to "informational support" demonstrated substantially higher positive SHAP values, while posts labels as "emotional support" yielded low SHAP values. This suggests that audiences engaging with mental health-related content prioritize actionable information over affective expression. This finding has important implications for understanding how social media platforms function as informal health information systems. The preference for informational over emotional support content indicates that users may be actively seeking mental health guidance and practical coping strategies through these platforms. This behavior aligns with emerging research on digital health-seeking and suggests opportunities for future investigation into how AI-driven content moderation and recommendation systems might be audited or designed to support mental health information access while maintaining safety and accuracy standards. Results in the visual analysis corroborate these findings. 

Additionally, videos appearing ``real'' (authentic footage) showed the strongest positive effect (Mean SHAP $\approx 0.5$), followed by professional-appearing production quality, wide shots, and visible text overlays (Mean SHAP $\approx 0.3$--$0.4$). Conversely, special effects showed the most dramatically negative contribution (Mean SHAP $\approx -1.3$). Split-screen formats, homemade appearance, AI-generated content, and selfie framing all showed modest negative effects. Not only do audiences seem to favor informative content, but they also have preferences for high perceived authenticity and visual production polish in mental health content.

\subsection{Cross-Modal Interactions}
Through modeling with XGBoost, we find that the interaction effects between visual and textual modalities reveal that differing levels of feature interactions drive engagement synergistically while others reduce engagement. 

Figure 4a illustrates how visual meme formatting and textual humor interact to shape engagement. The interaction exhibits a threshold effect: when the probability of meme formatting exceeds 50\%, the presence of humor in captions significantly amplifies engagement, as indicated by positive SHAP interaction values. This synergy is strongest when both features are highly probable. Below the 50\% meme probability threshold, humor shows no discernible effect on engagement, suggesting that ambiguous or weak meme formatting fails to activate audience expectations for humorous content. This uni-directional pattern indicates that meme formats function as a visual cue that primes audiences to reward humor, but this only occurs when the meme format is sufficiently recognizable. 

Figure 4b, showing the interaction between number of people present in image and textual probability of interpersonal relationship, demonstrates the opposite effect of Figure 4a. Here, divergence occurs on the left side of the graph. When videos contain no visible humans, discussions of interpersonal relationships increase viewership, as shown by positive SHAP interaction values. However, when one or more people appear in the visual frame, interpersonal relationship themes have no discernible effect on engagement. This  suggests that non-literal representations of social connection may enhance viewership. 

\begin{figure}[H]
    \centering
    \begin{subfigure}[t]{0.46\textwidth}
        \centering
        \includegraphics[width=\textwidth]{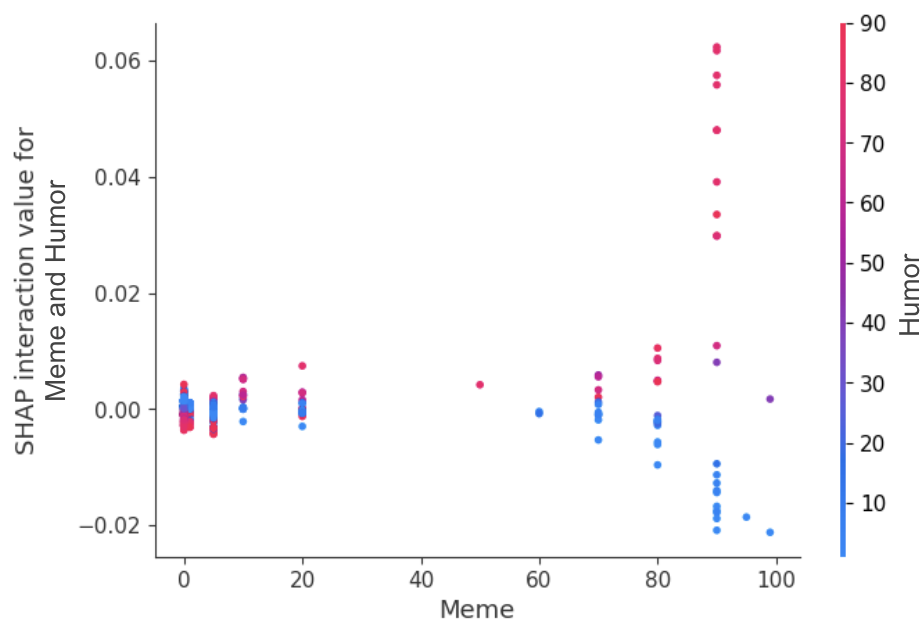} 
        \label{fig:int1}
        \caption{a) SHAP interaction between Meme format and caption humor}
    \end{subfigure}

    \begin{subfigure}[t]{0.46\textwidth}
        \centering
        \includegraphics[width=\textwidth]{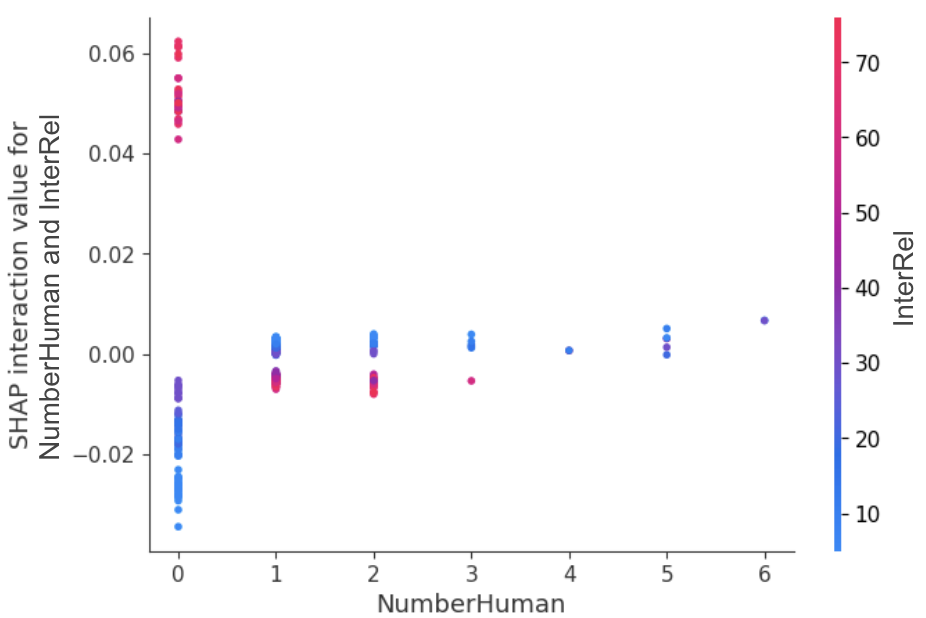} 
        \label{fig:int2}
        \caption{b) SHAP interaction between Number of People in frame and Interpersonal Relationship narratives in caption}
    \end{subfigure}

    \begin{subfigure}[t]{0.46\textwidth}
        \centering
        \includegraphics[width=\textwidth]{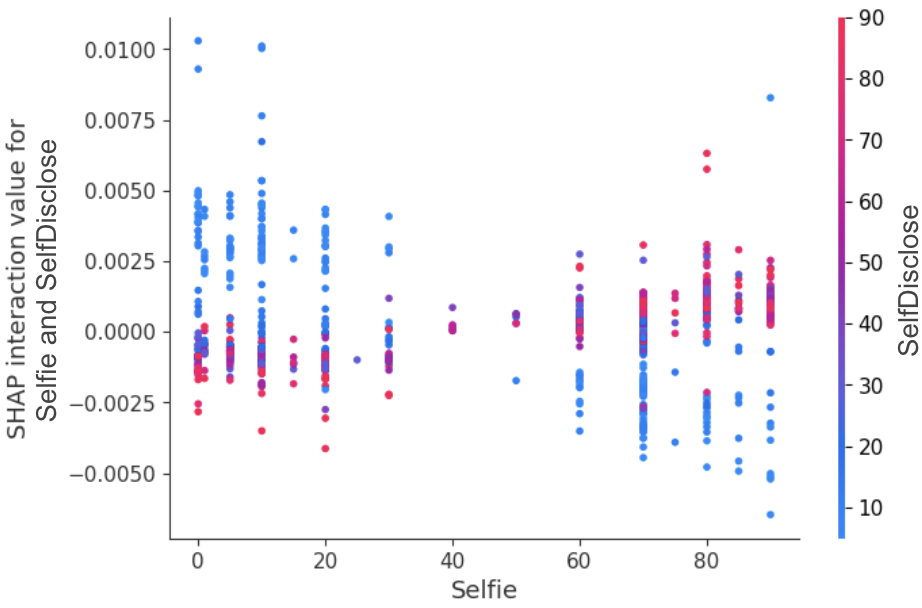}
        \label{fig:int3}
        \caption{c) SHAP interaction between Selfie visual format and Self Disclosure narrative in caption}
    \end{subfigure}
    
    \caption{SHAP interaction plots illustrating cross-modal feature effects on viewership. Y-axis shows SHAP Values. X-axis shows one feature's magnitude. 
    Blue represents low feature magnitude and red represents high magnitude.}
    \label{fig:1}
\end{figure}

Lastly, Figure 4c, illustrates how selfie framing and self-disclosure interact to shape viewership through a threshold-dependent relationship. When selfie probability is low (left side of the graph), high self-disclosure (red/pink points) is associated with negative SHAP interaction values, indicating reduced viewership. Conversely, when selfie probability exceeds approximately 60 (right side of the graph), high self-disclosure (red/pink points) shifts to positive SHAP interaction values, indicating increased viewership. This pattern reveals that self-disclosure narratives require visual self-presentation to drive viewership effectively. Personal storytelling demands visual self-exposure to signal authenticity and build virtual connection.

\begin{figure}[h!]
    \includegraphics[width=0.47\textwidth]{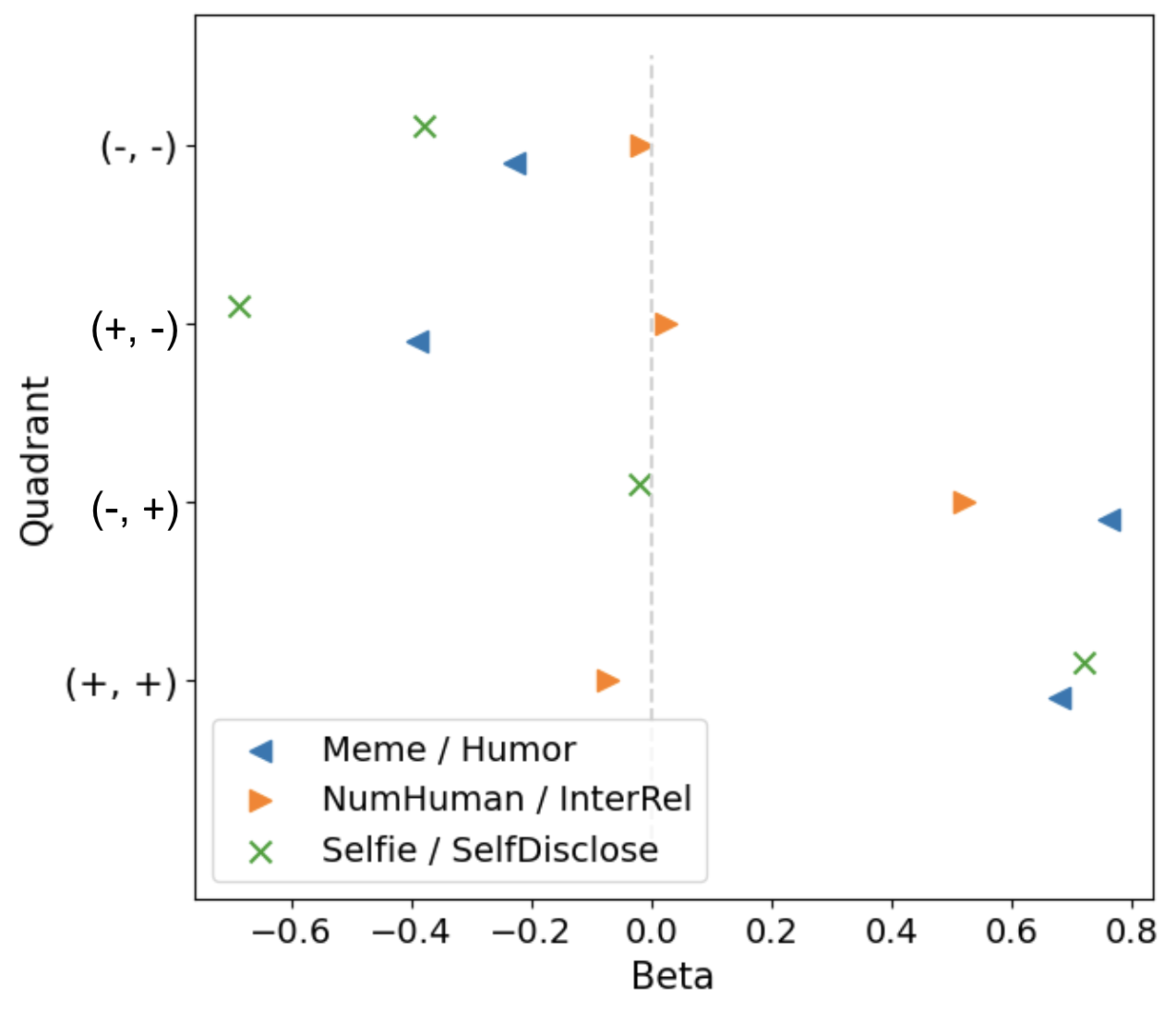}
    \caption{Piecewise regression coefficients from SHAP interaction values across quadrants. Each interaction pair is represented by a distinct marker. Four quadrants represent different combinations of feature values: (-, -) indicates both features below median threshold; (+, +) indicates both above threshold; (-, +) and (+, -) represent mixed conditions where one feature is above and one below.}
    \label{fig: quadrant}
\end{figure}

Similar to the beeswarm plot in Figure~\ref{fig:shap_sentiment}a, it is difficult to precisely interpret the SHAP interaction plots. To mathematically model how cross-modal interactions vary depending on feature intensity, we conducted piecewise regression analysis on SHAP interaction values across four quadrants using feature medians as thresholds. Figure~\ref{fig: quadrant} presents regression coefficients (beta values) for the three interaction pairs. 

The meme / humor interaction (blue triangles) shows clear synergy in the (+, +) and (-, +) quadrants. This indicates that high probability of humor in text correlates with an increase in viewership of the video irrespective of the meme format. Quadrants (-, -) and (+, -) show a moderate decrease in viewership. The number of humans / interpersonal relationships interaction (orange triangles) exhibits a different pattern: as noted previously, positive effects emerge when humans are absent but relationship discourse is present (-, + quadrant: $\beta \approx 0.3$), and all other quadrants have nearly no effect on viewership. This asymmetry suggests audiences may engage more readily with abstract relationship representations---text screenshots, empty spaces, metaphorical imagery---allowing viewers to project their own experiences onto ambiguous visual content. The selfie / self-disclosure interaction (green $\times$) demonstrates the most pronounced dependency: strongly negative when both features are absent (-, - quadrant: $\beta \approx -0.6$) and negative when self-disclosure occurs without visual self-presentation (-, + quadrant: $\beta \approx -0.3$), but robustly positive when both are present (+, + quadrant: $\beta \approx 0.7$).

The piecewise regression approach reveals interaction structures that would be obscured by linear modeling, providing more nuanced understanding of how multimodal features shape attention dynamics.

\subsection{Unimodal Measures are Inadequate in Multimodal Contexts}
Lastly, we construct an example where unimodal measurements lack nuance. A common assumption of existing social science research is that coarse features–typically measured through textual lexicon–serve as reliable and effective proxies for making inferences on attention dynamics. Our findings challenge this assumption using sentiment, a common predictive variable for social media engagement, to show that affective measures can diverge substantially across modalities within the same content and textual sentiment fail to capture nuance in negative emotion across domains. 

Figure~\ref{fig:importance} shows how facial emotions (as measured by Pyfeat) compare with textual sentiments from post caption and transcript sentiments. Since facial emotions are probabilities, points represent the average of emotions multiplied by the sentiment~\cite{wei2025faces}.

Figure~\ref{fig:importance} reveals inconsistent trends in the alignment between text sentiment, speech sentiment, and visual affect. Overall, captions are markedly flatter and more muted, with weighted average scores clustered in a narrow low range ($\approx 0.05$--$0.10$) across all emotion categories. In contrast, transcripts exhibit substantially higher and more differentiated sentiment weights ($\approx 0.17$--$0.29$), indicating stronger emotional expression in spoken content. Consequently, caption-based sentiment alone may systematically underestimate both emotional strength and emotional diversity relative to transcript-based analysis. The only consistent pattern observed across caption- and transcript-based sentiment is that sadness corresponds to the lowest sentiment in both written and spoken modalities. 


Additionally, comparing our findings to a different domain of discourse reveals contextual variations. Interestingly, recent work in political communication showed that the facial emotion of anger is correlated to the most negative textual sentiment~\citep{wei2025faces}. In mental health, sadness seemed to be most correlated with a negative sentiment. 
As Leo Tolstoy famously wrote in Anna Karenina (1877), ``Happy families are all alike; every unhappy family is unhappy in its own way.'' Similarly, the manifestation of negativity may be domain specific, with anger in poltics  sadness in mental health.

\begin{figure}[H]
    \includegraphics[width=0.48\textwidth]{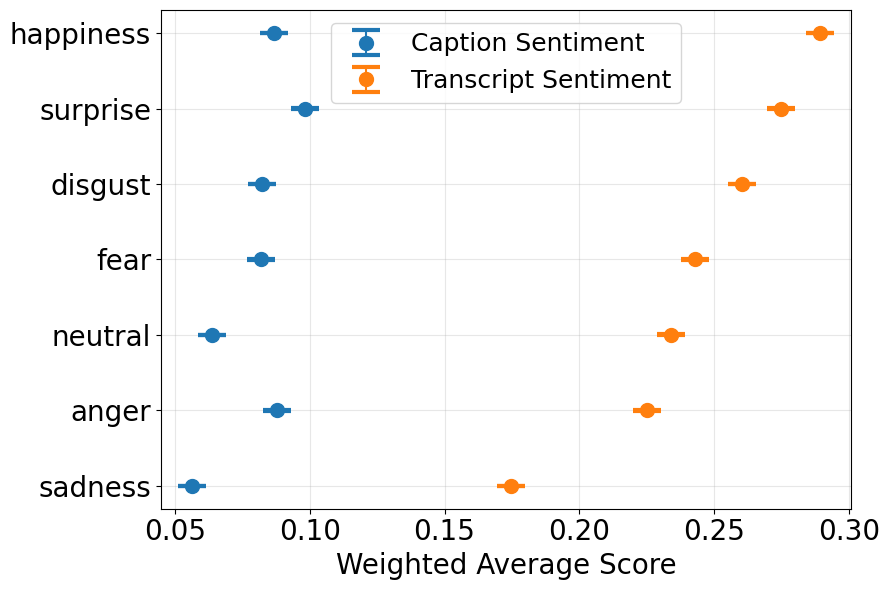}
    \caption{Comparison of Facial Emotion Scores with 95\% Confidence Intervals against Caption and Transcript Text Sentiment}
    \label{fig:importance}
\end{figure}



\section{Discussion}

\subsection{Methodological Contributions}

Multimodal communication on short-form video platforms presents computational social scientists with a fundamental challenge: how to systematically extract, integrate, and interpret heterogeneous features across text, visual, and audio modalities at scale while preserving interpretability. This study addresses that challenge by introducing a methodological framework that combines automated probabilistic feature extraction with Shapley value-based attribution, enabling researchers to quantify both individual feature contributions and cross-modal interactions in a unified explanatory framework.

Our approach makes three integrated methodological advances that together transform opaque multimodal content into interpretable, theory-grounded variables. First, zero-shot probabilistic classification enables theory-driven feature extraction without fine-tuning, representing discourse categories and visual attributes as continuous probability distributions (0-100) rather than discrete labels. This preserves measurement uncertainty, a theoretical central value, while producing features directly comparable across modalities. 

Second, we extend SHAP interpretability through feature-weighted aggregation ($\beta_{SHAP}$), consolidating point-level attributions into regression-style coefficients that social scientists can interpret as clarity in direction of impact, while respecting the nonlinear complexity of gradient-boosting trees. 

Third, piecewise regression on SHAP interaction values reveals threshold-dependent cross-modal synergies by partitioning feature space at median thresholds and estimating separate effects within each quadrant. This formalizes when modalities must coordinate versus when they operate independently. 

Applying this framework to 162,965 TikTok videos about social anxiety disorder demonstrates its empirical value. We show that sentiment diverges systematically across modalities, with facial expressions predicting engagement while textual sentiment produces noisy signals—evidence that single-modality proxies introduce systematic bias. Cross-modal interactions exhibit clear threshold effects: meme formats synergize with humor only when both are strongly present and selfie framing amplifies self-disclosure. These threshold-dependent patterns would be impossible to detect through manual coding at scale or through learned embeddings without interpretable attribution.

The framework generalizes beyond mental health communication through its modular architecture. Researchers can substitute domain-appropriate classification schemes (political frames, misinformation indicators) while maintaining the same analytical pipeline.

\subsection{Mental Health Communication}
In addition to methodological contributions, our findings reveal several substantive patterns about how mental health content functions in algorithmically mediated environments, with implications for both health communication research and platform design.

In affective analysis, the primacy of facial expressions over textual sentiment suggests visual self-presentation signals authenticity more effectively than text. Positive facial expressions, particularly happiness and neutral affect, predicted higher potentially because they signal approachability and emotional safety to audiences navigating sensitive content. This aligns with research on nonverbal communication showing that faces serve as privileged channels for affective authenticity~\citep{slepian2019facial}. This also has implications for peer support interventions and mental health advocacy campaigns, suggesting that visual warmth may be more critical in engaging audiences than text in multimodal contexts.

The preference for informational over emotional support indicates TikTok functions as an informal health information system rather than a traditional support group. Content labeled as providing informational support demonstrated higher viewership than content offering emotional validation or empathy. This pattern suggests users actively seek actionable guidance rather than affective connection, treating short-form video platforms as educational resources for mental health literacy. This behavior has important implications for understanding digital health-seeking and raises questions about information quality, accuracy, and potential harms. If audiences prioritize informational content from peer creators over emotional support, platforms should consider how recommendation algorithms might inadvertently amplify health misinformation or unqualified advice while suppressing professionally vetted resources.

\subsection{Limitations and Future Work}

Our study faces three limitations that point to methodological refinements. First, TikTok's practice of overlaying music with speech creates noise in audio attribution, as our pipeline does not separate these sources. Acoustic features thus capture composite signals rather than isolating speech or music independently—audio source separation techniques would enable more precise attribution. Second, our focus on TikTok limits generalizability, as algorithmic recommendation systems, user demographics, and interface affordances differ across platforms like Instagram Reels and YouTube Shorts. Cross-platform comparative studies would distinguish universal multimodal principles from platform-specific effects. Third, treating videos as static snapshots ignores how multimodal composition evolves temporally within content. Future work incorporating temporal feature extraction would reveal narrative dynamics and attention retention patterns throughout video duration.

The pipeline's modular architecture facilitates these extensions while supporting adaptation to other domains. Researchers can substitute classification schemes (political frames, misinformation indicators), outcome variables (shares, comments, conversion), and contextual features (creator demographics, posting time) to address questions in political communication, public health campaigns, or commercial content. Just as negative sentiment present differentially with facial affect, the dynamics of cross-modal interaction may yield fruitful, domain-specific outcomes. 

\section{Conclusion}



As computational social science increasingly grapples with rich, unstructured data from platforms where meaning-making operates multimodally, the methodological tools developed here become essential infrastructure. This study introduces a framework that combines automated probabilistic feature extraction with Shapley value-based attribution, enabling researchers to quantify individual feature contributions and cross-modal interactions while preserving interpretability. 
Our contribution, and hope, is that this helps envelop the necessary conditions for computation social science: analysis at scale, with theoretical grounding, and with interpretability sufficient for social scientific inference.

\bibliography{aaai2026}

\newpage

\section{Appendix}

\subsection{Textual/Transcript Classification Prompt}

\begin{tcolorbox}[
    enhanced,
    breakable,
    title=System prompt,
    colback=gray!10,
    colframe=gray!70,
    fonttitle=\bfseries,
    boxrule=0.8pt,
    arc=2pt,
    left=6pt,right=6pt,top=6pt,bottom=6pt
]

You are a public health researcher analyzing mental health discourse in social media post texts. Your task is to evaluate the following text: 

The following formatting notes are extremely important to follow exactly correctly:

Please give probabilities as percentage likelihood (i.e. 1\% if very unlikely and 99\% if extremely likely)

Corresponding to the questions below, you will need to output a JSON object. Return the structured JSON only, with no additional text, descriptions, or explanations. 

Mental health topics 

Please provide the probability that the text’s content relates to coping strategies?

Please provide the probability that the text’s content relates to communication in social interactions? 

Please provide the probability that the text’s content relates to interpersonal relationships?

Please provide the probability that the text’s content relates to self-growth?
Please provide the probability that the text’s content relates to situational stressors?

Please provide the probability that the text’s content contains the use of humor? 

Types of Social Support

Please provide the probability that the text’s content relates to emotional support (e.g., expressions of empathy, love, trust and caring)?

Please provide the probability that the text’s content relates to instrumental support (e.g., tangible aid and service)?

Please provide the probability that the text’s content relates to informational support (e.g., advice, suggestions, and information)?

Please provide the probability that the text’s content relates to appraisal support (e.g., information that is useful for self-evaluation)?

Based on the content, determine whether the text is seeking information, providing information, both, neither or unclear. Return this as a string: `"seeking"`, `"providing"`, `"both"`, `"neither"`, or `"unclear"`.

Content Themes

Please provide the probability that the text’s content relates to political issues.

Please also provide the probability that the text’s content relates to healthcare. 

Please also provide the probability that the text’s content relates to foreign affairs (e.g., foreign policy, war, global leaders). 

Please also provide the probability that the text’s content relates to economic issues. 

Please also provide the probability that the text’s content relates to climate issues. 

Please also provide the probability that the text’s content relates to COVID-19. 
Please also provide the probability that the text’s content relates to immigration. 

Please also provide the probability that the text’s content relates to crime or law enforcement. 

Please also provide the probability that the text’s content relates to technology. 

Please also provide the probability that the text’s content relates to large language models or AI chatbots. 

If large language models or AI chatbots are mentioned, also describe the stance as "positive", "neutral", "negative", or "unclear".

Please also provide the probability that the text’s content engages in advocacy, such as promoting a cause or calling for social or political change?

Please also provide the probability that the text’s content aims to raise awareness, such as by educating others about a specific issue or condition?

Please also provide the probability that the text’s content includes a call to action, such as encouraging the audience to take a specific step (e.g., donate, reach out, seek help)?

Please also provide the probability that the text’s content involves self-disclosure through personal narrative, such as sharing one's own experiences or mental health journey?

If the text mentions any mental health conditions (e.g., depression, anxiety, PTSD), list all mentioned conditions as strings.

If the text mentions other social media profiles (e.g., @username), list all usernames mentioned.

If the text references social media platforms other than TikTok (e.g., Instagram, Twitter), list all platforms referenced.

\end{tcolorbox}

\subsection{Visual Classification Prompt}

\begin{tcolorbox}[
    enhanced,
    breakable,
    title=System prompt,
    colback=gray!10,
    colframe=gray!70,
    fonttitle=\bfseries,
    boxrule=0.8pt,
    arc=2pt,
    left=6pt,right=6pt,top=6pt,bottom=6pt
]

You are a public health researcher analyzing mental health discourse in social media post visuals. Your task is to evaluate the following image. 

The following formatting notes are extremely important to follow exactly correctly:

If an image includes any text, please extract it all into a string. 

Please give probabilities as percentage likelihood (i.e. 1\% if very unlikely and 99\% if extremely likely)

Corresponding to the questions below, you will need to output a JSON object. Return the structured JSON only, with no additional text, descriptions, or explanations. 

Basic Identification

Describe this image in 200 words or less. Return as string. 

State whether the frame is most likely a real photo or AI-generated.

Please provide the probability that the frame contains text (label as Text:).

Please provide the probability that the frame uses special effects (label as SpecialEffects:).

Textual content: 

If the frame includes any text, please extract it all into a string (label as TextContent:).

If it contains text, what language is it (label as Language:)? If there are multiple languages, please include them as multiple fields.
If the frame includes any hashtags, please extract it all into a string (label as Hashtags:). 

If the frame includes any locations, please extract it all into a string (label as Location:). 

If the frame includes any date, please extract it all into a string (label as Date:). 

If the frame includes any link or QR code, please extract the link all into a string (label as Link:). 

If the frame includes any song names or music, please extract it all into a string (label as Music:). 

Visual content 

If the image contains an individual, what objects are the person interacting with directly? List all objects as strings. If there are multiple individuals, list all (label as RelevantObjects:).

Do you recognize any individuals or public figures from the frame? If so, who are they? (label as KnownIndividuals:)

Where is the setting of the location (e.g., bedroom, car, park, office, etc.) (label as Location:)?

Please provide the probability that the style of video shot is a close-up (label as CloseUp:).

Please provide the probability that the style of video shot is a full shot (label as FullShot:).

Please provide the probability that the style of video shot is a two shot (label as Two Shot:).

Please provide the probability that the style of video shot is a point-of-view (label as POV:).

Please provide the probability that the style of video shot is a Wide (label as Wide:).

Please provide the probability that the frame is split into segments (label as Split:).

How many segments is the frame split into? (label as Segments:).

How are the segments split? Vertically, horizontally or other? (label as Orientation:).

Please provide the probability that the recording is a Selfie, i.e., very close-up and clearly hand-held (label as Selfie:)

Please provide the probability that the recording is Homemade, i.e., non-professional and clearly not hand-held, such as footage filmed on a tripod (label as Homemade:).

Please provide the probability that the type of recording is professional (label as Professional:).

Please provide the probability that the frame depicts a meme (label as Meme:)?

What cartoon or anime source is this meme from (Family Guys, The Simpsons..etc) (label as MemeSource:)?

Please provide the probability that the frame is a drawing/illustration? (label as Illustration:)

Please provide the probability that the frame shows someone dancing (label as Dance:). 

Please provide the probability that the frame shows someone singing (label as Sing:).

\end{tcolorbox}

\subsection{Zero-shot Labeling Validation}

\begin{table}[ht]
\centering
\caption{Percent Validity of Caption Labeling by Category}
\label{tab:valid_by_cat_percent}
\begin{tabular}{l c}
\hline
\textbf{Category} & \textbf{Validity} \\
\hline
Appraisal Support & 75\% \\
Coping Strategies & 90\% \\
Emotional Support & 90\% \\
Healthcare & 100\% \\
Humor & 95\% \\
Informational Support & 90\% \\
Interpersonal Relationships & 100\% \\
Political Issues & 100\% \\
Self-Growth & 95\% \\
Situational Stressors & 65\% \\
\hline
\textbf{Overall} & \textbf{90\%} \\
\hline
\end{tabular}
\end{table}

\begin{table}[ht]
\centering
\caption{Percent Validity of Transcript Labeling by Category}
\label{tab:valid_by_cat}
\begin{tabular}{l c}
\hline
\textbf{Category} & \textbf{Validity} \\
\hline
Appraisal Support & 85\% \\
Communication Probability & 75\% \\
Coping Strategies & 90\% \\
Emotional Support & 70\% \\
Healthcare & 90\% \\
Humor & 85\% \\
Informational Support & 95\% \\
Instrumental Support & 85\% \\
Self-Disclosure Narrative & 100\% \\
Self-Growth & 90\% \\
\hline
\textbf{Overall} & \textbf{86.5\%} \\
\hline
\end{tabular}
\end{table}

\begin{table}[H]
\centering
\caption{Percent Validity of Image Labeling by Category}
\label{tab:valid_by_image_type}
\begin{tabular}{l c}
\hline
\textbf{Category} & \textbf{Validity} \\
\hline
AI & 60\% \\
Full Shot & 75\% \\
Homemade & 100\% \\
Professional & 70\% \\
Real & 100\% \\
Selfie & 75\% \\
Special Effects & 75\% \\
Split & 100\% \\
Text & 85\% \\
Wide & 95\% \\
\hline
\textbf{Overall} & \textbf{83.5\%} \\
\hline
\end{tabular}
\end{table}

\subsection{Full Shap Beeswarm Plots}

\begin{figure}[H]
    \centering
    \begin{subfigure}[t]{0.48\textwidth}
        \centering
        \includegraphics[width=\textwidth]{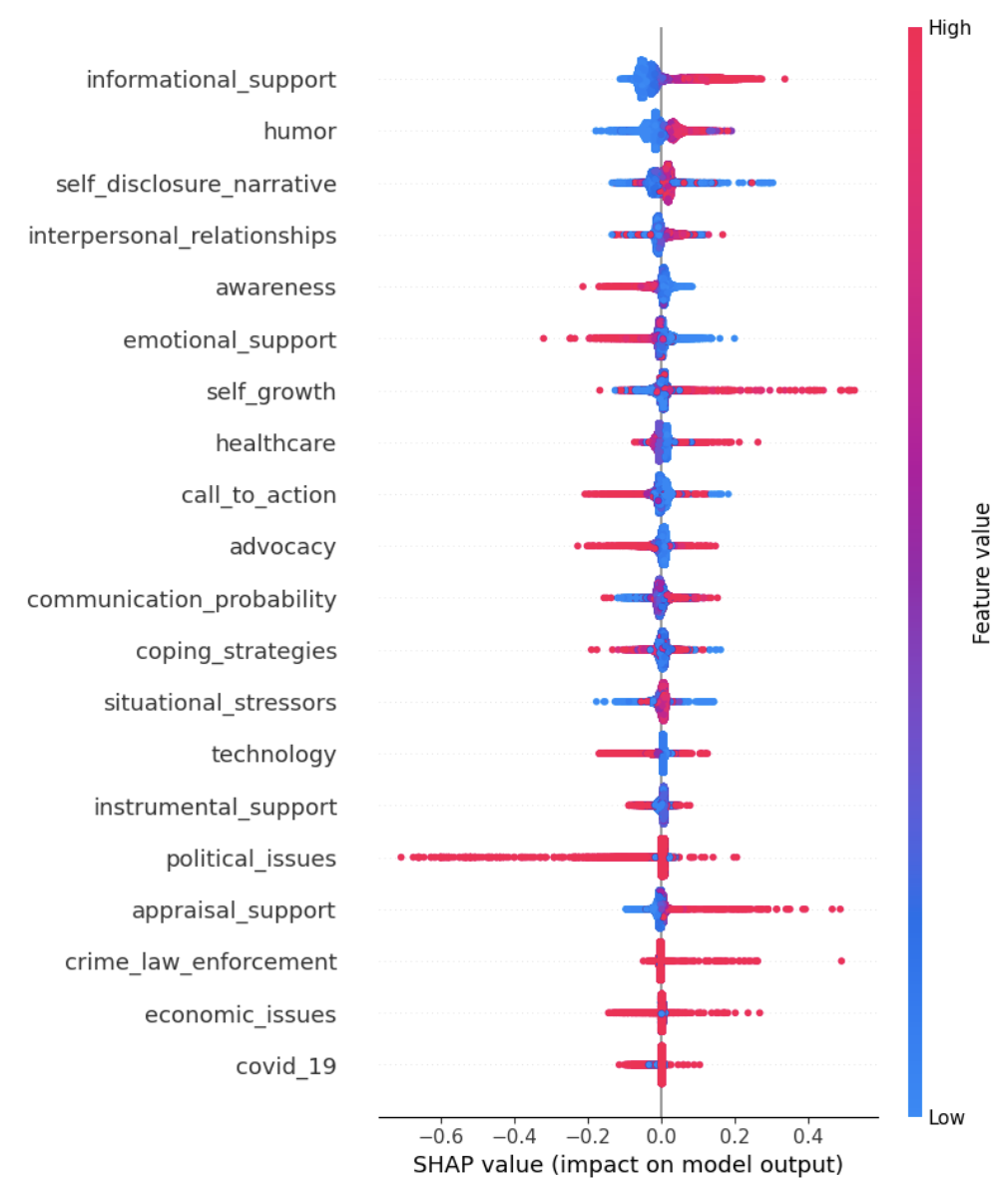}
        \caption{SHAP value distributions for probabilistic textual features.}
        \label{fig:2a}
    \end{subfigure}
    
    \vspace{0.5cm} 

\end{figure}

\begin{figure}[H]
    \ContinuedFloat
    
    \begin{subfigure}[t]{0.48\textwidth}
        \centering
        \includegraphics[width=\textwidth]{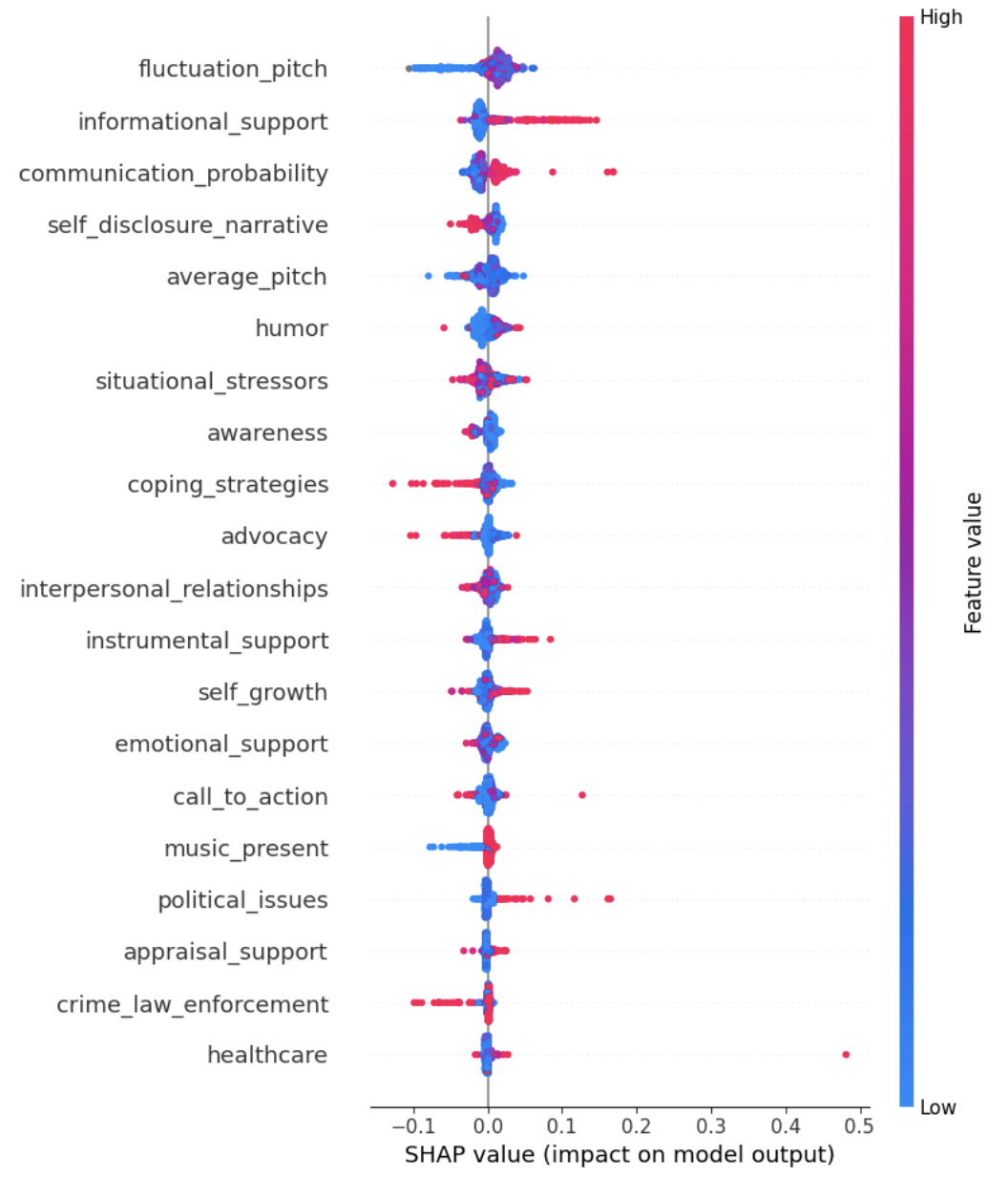} 
        \caption{SHAP value distributions for probabilistic auditory features.}
        \label{fig:2b}
    
    \end{subfigure}
    
        \begin{subfigure}[H]{0.48\textwidth}
        \centering
        \includegraphics[width=\textwidth]{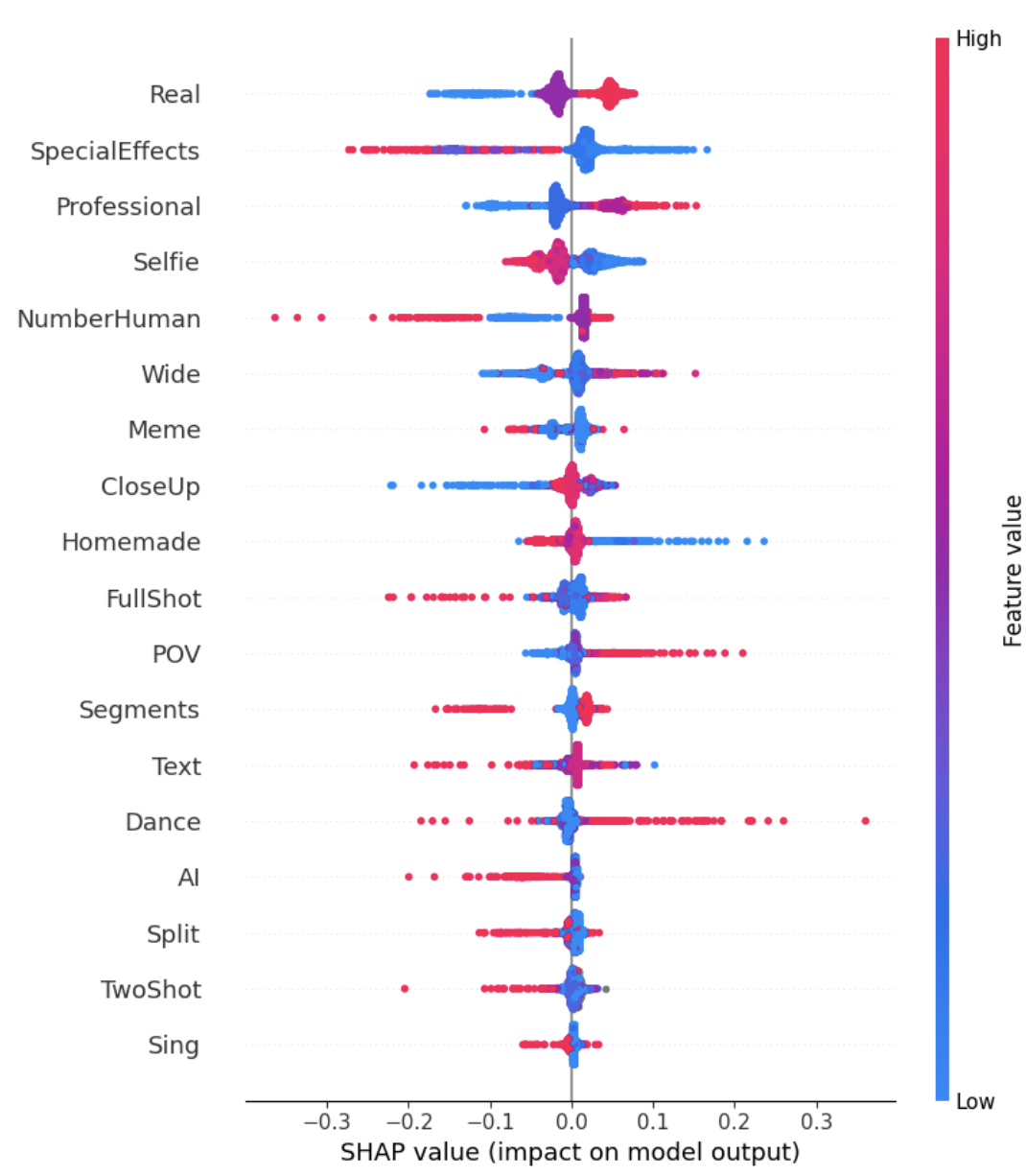} 
        \caption{SHAP value distributions for probabilistic visual features.}
        \label{fig:2c}
    \end{subfigure}

    \caption{SHAP Summary Plot of textual, visual, and auditory features. Colors represent the feature magnitude,
where blue represents low magnitude and red represents
high magnitude.}
    \label{fig:1}
\end{figure}

\subsection{Regression Plots}

\begin{figure}[H]
    \centering
    \begin{subfigure}[t]{0.45\textwidth}
        \centering
        \includegraphics[width=\textwidth]{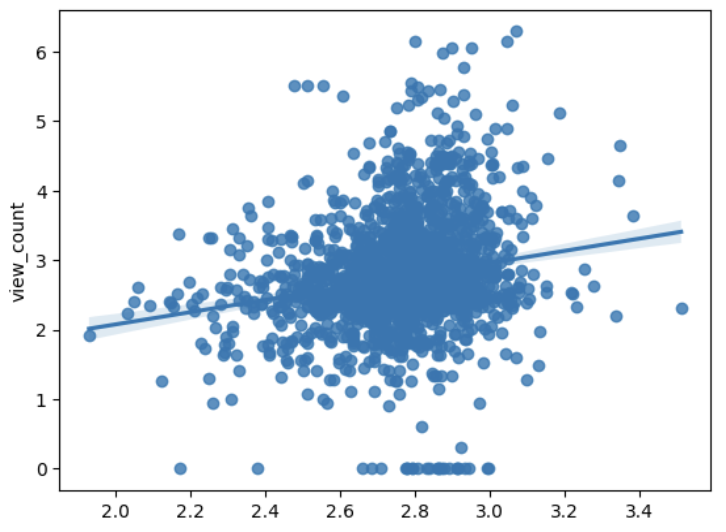}
        \caption{Scatter plot with fitted regression line showing the relationship between predicted viewership using textual features (x-axis) and actual view counts (y-axis).}
        \label{fig:2a}
    \end{subfigure}
    
    \vspace{0.3cm} 

    \begin{subfigure}[t]{0.45\textwidth}
        \centering
        \includegraphics[width=\textwidth]{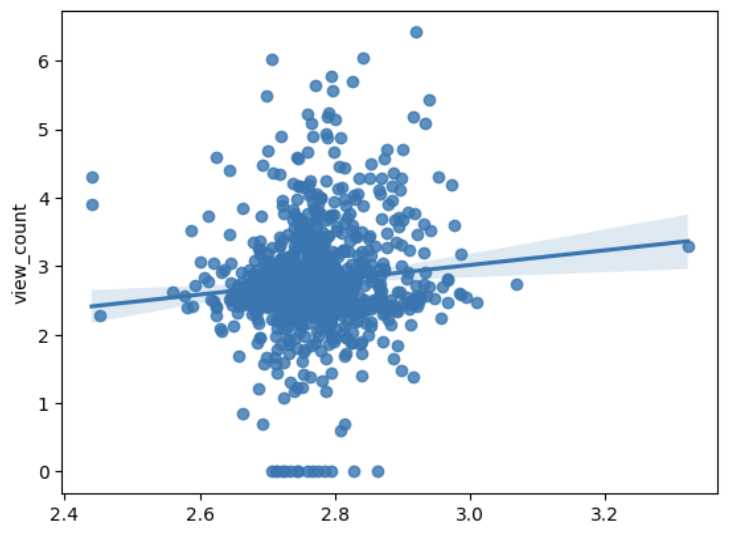} 
        \caption{Scatter plot with fitted regression line showing the relationship between predicted viewership using visual features (x-axis) and actual view counts (y-axis).}
        \label{fig:2b}
    \end{subfigure}

    \vspace{0.3cm} 

    \begin{subfigure}[t]{0.45\textwidth}
        \centering
        \includegraphics[width=\textwidth]{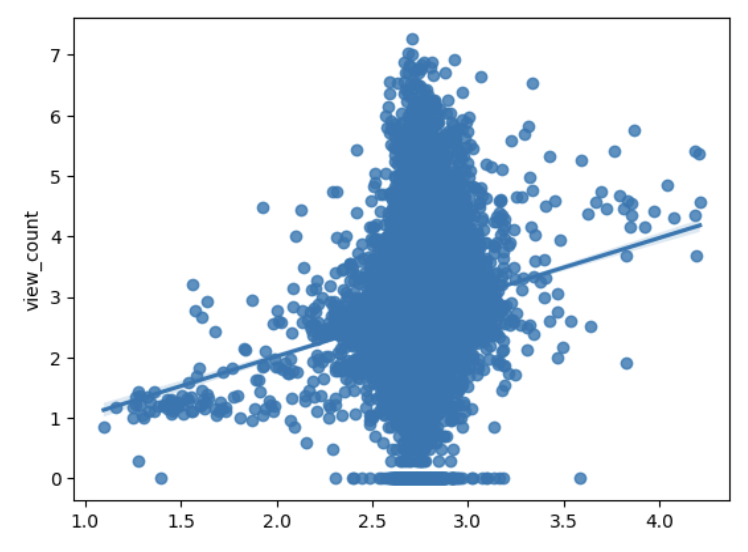} 
        \caption{Scatter plot with fitted regression line showing the relationship between predicted viewership using auditory features (x-axis) and actual view counts (y-axis).}
        \label{fig:2c}
    \end{subfigure}

    \caption{Regression plots for textual, visual, and auditory features.}
    \label{fig:1}
\end{figure}

\end{document}